# Interdisciplinary Research Methodologies in Engineering Education Research

David Reynolds [a] & Nicholas Dacre [b]

[a] WMG, University of Warwick, Coventry, UK david.reynolds@warwick.ac.uk
[b] University of Southampton Business School, Southampton, UK nicholas.dacre@southampton.ac.uk

## Abstract

As Engineering Education Research (EER) develops as a discipline it is necessary for EER scholars to contribute to the development of learning theory rather than simply being informed by it. It has been suggested that to do this effectively will require partnerships between Engineering scholars and psychologists, education researchers, including other social scientists. The formation of such partnerships is particularly important when considering the introduction of business-related skills into engineering curriculum designed to prepare 21st Century Engineering Students for workplace challenges. In order to encourage scholars beyond Engineering to engage with EER, it is necessary to provide an introduction to the complexities of EER.

With this aim in mind, this paper provides an outline review of what is considered 'rigorous' research from an EER perspective as well as highlighting some of the core methodological traditions of EER. The paper aims to facilitate further discussion between EER scholars and researchers from other disciplines, ultimately leading to future collaboration on innovative and rigorous EER.

## Introduction

There is a perceived "skills mismatch between what engineering graduates possess and what is demanded by industry and potential employers" (Bubou et. al., 2017). Therefore, to prepare 21st Century Engineering students for the reality of the workplace, an Engineering curriculum should also include business-related skills such as Marketing (Rammant, 1988), Project Management (Dacre et. al., 2019; Pons, 2015) and other 'soft' skills (Wilson & Marnewick, 2018) or "professional competences" (Carthy et.al., 2018). This provides a particular challenge for academics from within these business-related disciplines who may wish to conduct Engineering Education Research (EER). As is the case with many EER scholars, many of these academics will be under-resourced and will be conducting education research part-time (Shawcross & Ridgman, 2013). Hence, they are likely to focus on EER areas in which they have an intrinsic interest (Nyamapfene & Williams, 2017) and due to familiarity, they may simply apply the traditions and approaches of their 'home' discipline rather than those of EER (Borrego & Streveler, 2015). If the traditions of their 'home' discipline are significantly different to the traditions of EER this may lead to issues when attempting to disseminate any findings in more traditional EER outlets.

Despite a rapidly growing body of literature, EER is still considered an emerging field of enquiry (Borrego & Streveler, 2015; Liu, 2019). Unsurprisingly, in the early stages of emergence, there is significant debate regarding the many possible epistemological, ontological and methodological approaches that could be applied to EER (Borrego et al., 2009); Borrego & Bernhard, 2011). It has been documented that there are often epistemological tensions among EER scholars (Cicek & Friesen, 2018) and these are likely to be further strained by the introduction of academics from different disciplines and traditions.

It is beyond the scope of this paper to provide an in- depth analysis of the various approaches currently used in EER. Instead, it is the authors' intention to highlight some of the main frameworks and models applied to EER in order to encourage discussion about how to embed pedagogic research related to businessrelated skills within the burgeoning traditions of EER. The authors also hope that this paper will encourage the collaboration between engineering faculty and social scientists necessary to allow EER to contribute to learning theory (Streveler & Smith, 2006).

## What is Rigorous EER?

Just as Engineering is viewed as a scientific discipline, many definitions of rigour in EER can be traced back to definitions of rigour in scientific education research (Streveler and Smith, 2006). Moreover, there appears to be a general consensus among scholars of EER that science education research (including EER) should look beyond simply examining methods of improving the practice of teaching in class (Fensham, 2004; Lattuca & Litzinger, 2015). Instead, to be recognised as a discipline in its own right EER should also aspire to contribute to both theoretical and conceptual developments about how students learn Engineering (Streveler and Smith, 2006; Borrego and Streveler, 2015). Based on an initial review of the relevant literature, this means EER research should be:

1. Problem-led, hence requiring empirical investigation (Shavelson & Towne, 2002; Borrego and Bernhard, 2011; Bernhard & Baillie, 2016; Malmi, et al., 2018).

2. Informed by (and inform) both relevant educational theory and discipline specific theory (Shavelson & Towne, 2002; Streveler and Smith, 2006; Borrego and Bernhard, 2011; Bernhard & Baillie, 2016; Malmi, et al., 2018).

3. Method-led, meaning the methods used must be consistent and relevant to the question being investigated (Shavelson and Towne, 2002; Borrego and Bernhard, 2011; Bernhard & Baillie, 2016; Malmi, et al., 2018).

4. Systematic, explicit and provide a coherent and explicit chain of reasoning (Shavelson and Towne, 2002; Malmi, et al., 2018).

5. Presented in a way that allows it to be open to professional scrutiny and critique by both academics and practitioners (Shavelson & Towne, 2002; Borrego & Bernhard, 2011; Borrego & Streveler, 2015; Bernhard & Baillie, 2016; Malmi, et al., 2018)

As should be apparent from the above, despite its scientific origins, EER is more generally viewed as interdisciplinary in nature (Malmi et al., 2018). Hence while the above is an attempt to define what rigorous EER is, academics from business-related disciplines wishing to conduct EER would also benefit from a framework to assess if their research would be viewed as rigorous by the EER community. Recently, Borrego and Bernhard (2016) offered a "tentative quality criteria" for qualitative EER research.

This criteria was separated into three parts: Quality of the study in general; Quality of the Results; and Validity of the Results. They claim this set of criteria is consistent with other lists of criteria, including those from the Journal of Engineering Education and the European Journal of Engineering Education (Borrego and Bernhard, 2016). It could also be argued that these criteria should be applied to EER in general and are just as relevant to quantitative, constructive and mixed-method research.

Despite widespread agreement regarding the need for rigorous EER, what does appear to be open for debate is how "generalizable" the findings of EER need to be in order to be considered worthy/rigorous. Some academics argue that "generalizable" is an essential criteria of all scientific research and hence the same should apply to EER (Shavelson and Towne, 2002; Malmi, et al 2018; Streveler and Smith, 2006). However, others such as Bernhard & Baillie (2016) argue that EER is "situated in international and interdisciplinary contexts" and hence results may not be "generalizable/transferable to other contexts (disciplines and/or countries)". Regardless, it is clear that there is a need for academics wishing to conduct EER to make explicit the underlying epistemological and ontological perspectives of their research.

## What is Considered Appropriate Evidence in EER?

It has been suggested that the nature of higher education student experience is related to the methodologies employed by higher education researchers (Khan, 2015). Furthermore, in order for EER to be effective in identifying ways to improve engineering education it should be learnercentred or student-centred (Catalano & Catalano,1999) and requires "multiple epistemic frames" (Riley, 2014). In fact, in their study of 155 EER papers, Malmi et al. (2018) identified 128 different explanatory frameworks. Therefore, based on the work of Bubou et al. (2017), the purpose of this section is to introduce the three most popular traditions currently being applied in EER in order to give scholars who are new to this field a number of options from which to position their own research.

Bubou et al. (2017) identified three traditions within EER: Discipline-based education research (DBER); Scholarship of Teaching and Learning (SoTL) and Evidence-Based Teaching (EBT). The oldest of these traditions, the knowledge base of DBER, has been built in over 30 years (Bubou et al., 2017). This tradition tends to emphasise improvements in the practice of teaching, usually focussing on a specific topic. See also the work of Hutchings & Shulman (1999) into "effective teaching" and "scholarly teaching" (cited in Borrego & Strevler, 2015). Research based in this tradition may be viewed as "teacher-centric" rather than "student-centric" (Hamer, 2006; Pears et al., 2016) with an emphasis on identifying how to best 'transfer' teachers' knowledge to students.

This approach encourages the use of experimentation and comparative studies, using changes in student grades and attendance as evidence of change. This is not intended as a criticism, as research using this framework is clearly important for the development of teaching 'best practice' and DBER has been widely published in science research journals including proceedings from the National Academy of Science (Bubou et al., 2017).

In contrast, SoTL emphasises student-centrality, and encourages the systematic investigation of student learning as a concept (Bubou et al., 2017). SoTL research embraces discussion and critique beyond the classroom, but also tends to be topic-specific (Borrego & Streveler, 2015). As a result, researchers within the SoTL will often use discovery and reflection as sources of evidence (Bubou et al., 2017).

The last tradition, EBT, could be viewed as an attempt to bridge the gap between DBER and SoTL. Inspired by Evidence-Based Medicine, EBT encourages the the collection, analysis, and interpretation of information about students to inform teaching and learning for better outcomes for the education system as a whole (Bubou et al., 2017). See also "data-based decision making" in Education, such as the work of Škėrienė and Augustinienė (2018).

EBT is based on social constructivist learning theories, the science of learning (learning sciences), and teaching/learning styles (Bubou et al., 2017). As a result, EBT research tends to encourage the use of a wide variety of "emerging" (usually qualitative) methodologies including Case Study, Grounded Theory, Ethnography,

Action Research, Phenomenography, Discourse Analysis, and Narrative Analysis (Case and Light, 2011).

## Conclusion

The authors of this paper agree with Borrego et al. (2009) who state that "no particular method is privileged over any other". However, standards of rigour in EER must be maintained. Bernhard & Baillie (2016) warn that "An unhealthy overemphasis on either [problem-led or method-led research] can lead to a lack of quality".

This is particularly important when conducting interdisciplinary research. It is also important when engaging with models and theories which investigate factors that influence learning beyond what occurs in the classroom (Streveler and Smith, 2006). For example, investigations into encouraging wider student engagement and the development of engineering curriculum appropriate for the needs of both engineering students and the organisations which will eventually employ them (Lattuca and Litzinger, 2015).

Streveler and Smith (2006) argue that rigorous EER should contribute to learning theory, rather than merely being informed by it. However they also claim in order to achieve this aim, it is necessary to foster partnerships with "psychologists, education researchers, or other social scientists" (Streveler and Smith, 2006). We propose that in order to effectively introduce researchers from these disciplines to EER, it is necessary to highlight the underlying epistemic, ontological and methodological traditions (and debates) within this discipline.

It is our hope that this paper will serve as an introduction to some of the terminology, frameworks, and models in EER. For researchers seeking more detailed discussion about these traditions and recent developments within EER, we direct readers to the sources used within this paper, but most significantly the comprehensive Cambridge Handbook of Engineering Education Research (Johri and Olds, 2014).